\documentclass[pra,aps,amsfonts,amsmath,twocolumn,showpacs,floatfix]{revtex4}
\usepackage{bm}
\usepackage{amsfonts}
\usepackage{amsmath}
\usepackage{epsfig}
\newcommand{\eps}{\varepsilon}
\newcommand{\Ref}[1]{Ref.\@ \cite{#1}}
\newcommand{\Refs}[1]{Refs.\@ \cite{#1}}
\newcommand{\Eq}[1]{Eq.\@ (\ref{#1})}
\newcommand{\Eqs}[1]{Eqs.\@ (\ref{#1})}
\newcommand{\Sec}[1]{Sect.\@ \ref{#1}}
\newcommand{\Fig}[1]{Fig.\@ \ref{#1}}
\newcommand{\calH}{\mathcal{H}}
\newcommand{\calR}{\mathcal{R}}
\newcommand{\Ep}{E^\prime}

\newcommand{\np}{{n^\prime}}
\newcommand{\Eb}{\bar{E}}

\newcommand{\xip}{\xi^\prime}

\newcommand{\xib}{\bar{\xi}}
\newcommand{\calHtil}{\tilde{\mathcal{H}}}
\newcommand{\calRtil}{\tilde{\mathcal{R}}}
\newcommand{\Deltatil}{\tilde{\Delta}}
\newcommand{\htil}{\tilde{h}}
\newcommand{\kappatil}{\tilde{\kappa}}
\newcommand{\rhotil}{\tilde{\rho}}
\newcommand{\Av}{\bm{\mathrm{A}}}
\newcommand{\Bv}{\bm{\mathrm{B}}}
\newcommand{\ev}{\bm{\mathrm{e}}}
\newcommand{\jv}{\bm{\mathrm{j}}}
\newcommand{\pv}{\bm{\mathrm{p}}}
\newcommand{\rv}{\bm{\mathrm{r}}}
\newcommand{\sv}{\bm{\mathrm{s}}}
\newcommand{\vv}{\bm{\mathrm{v}}}
\newcommand{\nablav}{\bm{\nabla}}
\newcommand{\poisson}{\overset{\leftrightarrow}{\Lambda}}
\DeclareMathOperator{\arsinh}{arsinh}
\begin{document}
\title{Slow rotation of a superfluid trapped Fermi gas}
\author{Michael Urban}
\affiliation{Institut de Physique Nucl{\'e}aire, F-91406 Orsay
  C{\'e}dex, France}
\author{Peter Schuck}
\affiliation{Institut de Physique Nucl{\'e}aire, F-91406 Orsay
  C{\'e}dex, France}
\begin{abstract}
The moment of inertia, $\Theta$, is one of the possible observables
for the experimental determination whether a trapped Fermi system has
reached the BCS transition or not. In this article we investigate in
detail the temperature dependence of $\Theta$ below the critical
temperature $T_c$. Special care is taken to account for the small size
of the system, i.e., for the fact that the trapping frequency
$\hbar\omega$ is of the same order of magnitude as the gap
$\Delta$. It is shown that the usual transport approach, corresponding
to the leading order of an expansion in powers of $\hbar$, is not
accurate in this case. It turns out that $\Theta$ does not change
rapidly if $T$ becomes smaller than $T_c$, but it rather decreases
slowly. Qualitatively this behavior can be explained within the
two-fluid model, which again corresponds to the leading order in
$\hbar$. Quantitatively we find deviations from the two-fluid model
due to the small system size.
\end{abstract}
\maketitle
%
\section{Introduction}
%
Since the first observation of Bose-Einstein condensation of
magnetically trapped bosonic atoms \cite{Anderson,Davis,Bradley} it
has become clear that ultra-cold trapped atomic gases provide an
excellent tool to study quantum effects in systems which are almost
visible to the naked eye. For example, quantized vortices in the Bose
condensate were created by stirring the Bose condensate with the help
of a laser beam \cite{Madison}. Also the quantum pressure related to
the Pauli principle could be observed in gases of trapped fermionic
atoms \cite{Truscott,Schreck}, which proves that temperatures well
below the degeneracy temperature can be reached.

If it was possible to trap two spin states of a fermionic isotope with
attractive interaction, and to cool the system below the critical
temperature $T_c$, one could study the BCS transition to the
superfluid phase. Unlike the transition of a Bose gas to the
Bose-Einstein condensate, the BCS transition of a Fermi gas almost
does not change the density profile of the atomic cloud
\cite{Houbiers}. However, there are other observables which may allow
to distinguish between the normal-fluid and the superfluid phase. In a
preceding paper \cite{Farine} the moment of inertia was proposed,
since it is much smaller in the superfluid phase than in the
normal-fluid phase (see also \Ref{Zambelli}). Another observable
changing from one phase to the other are the frequencies of collective
modes \cite{Zambelli,Baranov,BruunMottelson}. For example, the
frequency of the so-called ``scissors mode'', an oscillation of the
symmetry axis of the cloud with respect to the symmetry axis of the
trap, is closely related to the moment of inertia
\cite{Farine}. Recently one more observable for the detection of the
BCS transition was proposed, namely the change of the deformation of
the cloud during the expansion of the system when the trapping
potential is switched off \cite{Menotti}.

The moment of inertia of a superfluid gas of trapped fermionic atoms
at zero temperature was evaluated for the first time in \Ref{Farine}
in close analogy to the calculation of the moment of inertia of
superfluid nuclei \cite{Durand}. This derivation was very similar to
the one given by Migdal more than 40 years ago \cite{Migdal}, except
that everything was reformulated in phase space in terms of Wigner
transforms. In the present article we will generalize the calculation
of \Ref{Farine} to the case of non-zero temperature. In addition, we
will give a derivation which further clarifies certain points which in
\Ref{Farine} may have been passed over rather quickly.

In addition to the temperature dependence of the moment of inertia, we
will address an interesting question which is relevant already at zero
temperature. In nuclear physics it is well known that the moment of
inertia of superfluid nuclei is much smaller than the rigid-body
value, but still higher than the value corresponding to a purely
irrotational motion, and that the currents in rotating nuclei have
both rotational and irrotational components \cite{Durand}. The same
behavior is found in trapped Fermi gases at zero temperature
\cite{Farine}. In contrast to this, the ordinary hydrodynamical or
transport equations for superfluids at zero temperature, which can be
derived from the $\hbar\to 0$ limit of the time-dependent
Hartree-Fock-Bogoliubov (TDHFB) equation
\cite{Betbeder,Serene,DiToro,Woelfle,Gulminelli} allow only for a
purely irrotational motion. We will work out this difference and
discuss the limits of validity of the hydrodynamical description.

The article is organized as follows: In \Sec{formalism} we give a
brief review of the formalism, mainly in order to recall some
definitions and to clarify our notation. In \Sec{linres} we derive the
expression for the density matrix of the slowly rotating system within
linear-response theory. This is the generalization of the calculation
of \Ref{Farine} to non-zero temperatures. In \Sec{transport} we again
derive the linear response of the density matrix, but now using the
leading order of the $\hbar$ expansion of the TDHFB equation. In
\Sec{results} we show numerical results for the moment of inertia
obtained within both formalisms as a function of temperature and
interprete the results and their differences. Finally, in
\Sec{summary}, we summarize and draw our conclusions.
%
\section{Brief review of the formalism}
%
\label{formalism}
Before considering the rotating superfluid trapped Fermi gas, we will
briefly review the equilibrium case. Our intention is to explain our
notation and conventions. Detailed discussions of the subject can be
found in many articles \cite{Houbiers,Farine,BruunCastin} and
textbooks \cite{RingSchuck}.

In this article we assume for simplicity that equal numbers of atoms
with two spin projections $\sigma = \uparrow,\downarrow$ are trapped
in a spin-independent harmonic potential
\begin{equation}
V_0(\rv) =
\sum_{i=xyz}\frac{m\omega_{0i}^2}{2}r_i^2\,.
\end{equation}
If the density of the trapped system is very low, the atom-atom
interaction can be approximated by a zero-range interaction with a
coupling constant $g$ proportional to the $s$-wave scattering
length. Due to the Pauli principle only atoms with opposite spin
projections can interact in this way. Under these assumptions the
hamiltonian takes the form
\begin{multline}
H = \int\! d^3 r \Big[\sum_{\sigma=\uparrow,\downarrow}
  \psi_\sigma^\dagger(\rv)
  \Big(-\frac{\hbar^2\nablav^2}{2m}+V_0(\rv)\Big)
  \psi_\sigma(\rv)\\
   -g \psi_\downarrow^\dagger(\rv)\psi_\uparrow^\dagger(\rv)
   \psi_\uparrow(\rv)\psi_\downarrow(\rv)\Big]\,.
\label{hamiltonian}
\end{multline}

The mean-field potential corresponding to this interaction reads
\begin{equation}
V(\rv) = V_0(\rv)-g \rho(\rv,\rv) = V_0(\rv)-g\rho(\rv)\,,
\label{vmeanfield}
\end{equation}
where we have used the following notation for the non-local density
matrix:
\begin{equation}
\rho(\rv,\rv^\prime) =
\langle\psi_\uparrow^\dagger(\rv^\prime)\psi_\uparrow(\rv)\rangle =
\langle\psi_\downarrow^\dagger(\rv^\prime)\psi_\downarrow(\rv)\rangle
\,.\label{rhocoo}
\end{equation}
(Note that with this definition the local part of the density matrix,
$\rho(\rv) \equiv \rho(\rv,\rv)$ corresponds to the density per spin
state.) In the presence of pairing correlations, the pairing gap is
given by the gap equation
\begin{equation}
\Delta(\rv) = g \kappa(\rv,\rv)
\label{gapeqcoo}
\end{equation}
where the pairing tensor has been defined as
\begin{equation}
\kappa(\rv,\rv^\prime) =
\langle\psi_\downarrow(\rv^\prime)\psi_\uparrow(\rv)\rangle\,.
\end{equation}
It will turn out that the self-consistent solution of \Eq{gapeqcoo} is
divergent as a consequence of the zero-range interaction. In the
literature several ways how to regularize this divergence can be found
\cite{Houbiers,Farine,BruunCastin,SadeMelo}, but in fact the technical
details of the solution of \Eq{gapeqcoo} are not important for our
purpose.

In order to write down the Hartree-Fock-Bogoliubov (HFB) equations,
which relate the density matrix $\rho$ and the pairing tensor
$\kappa$ to the potential $V$ and the gap $\Delta$, it is useful to
expand all quantities in a basis of single-particle wave functions
$\varphi_n(\rv)$, where $n$ represents all quantum numbers except
spin, i.e., for an arbitrary operator $A$:
\begin{equation}
A_{n\np} = \int\!d^3r\,d^3r^\prime\,\varphi_n^*(\rv)
  \varphi_\np(\rv^\prime)A(\rv,\rv^\prime)\,.
\end{equation}
Expressing the field operators $\psi_\sigma(\rv)$ and
$\psi^\dagger_\sigma(\rv)$ in terms of annihilation and creation
operators $a_{n\sigma}$ and $a^\dagger_{n\sigma}$, we recover the
usual definitions
\begin{gather}
\rho_{n\np} = \langle a^\dagger_{\np\uparrow}a_{n\uparrow}\rangle\,,\\
\kappa_{n\np} = \langle a_{\bar{n}^\prime\downarrow} a_{n\uparrow}\rangle\,.
\label{kappabas}
\end{gather}
The index $\bar{n}^\prime$ in \Eq{kappabas} denotes the time-reversed
state characterized by $\varphi_{\bar{n}^\prime}(\rv) =
\varphi_\np^*(\rv)$. We need also the matrix elements $h_{n\np}$ of
the grand-canonical (mean-field) single-particle hamiltonian (i.e., of
the single-particle hamiltonian minus the chemical potential $\mu$)
\begin{equation}
h = \frac{\pv^2}{2m}+V(\rv)-\mu\,,
\label{hsp}
\end{equation}
and the matrix elements $\Delta_{n\np}$ of the gap $\Delta$. For the
more general case that the hamiltonian is not time-reversal invariant,
we introduce the notation $\bar{A}_{n\np} =
A_{\bar{n}^\prime\bar{n}}$. If the matrices mentioned above are
combined as follows:
\begin{gather}
\calR = \Big(\begin{array}{cc}\rho& -\kappa\\
  -\kappa^\dagger&1-\bar{\rho}\end{array}\Big) \,,\\
\calH = \Big(\begin{array}{cc}h& \Delta\\
  \Delta^\dagger&-\bar{h}\end{array}\Big)\,,
\end{gather}
the HFB equations \cite{RingSchuck,Valatin} can be written in the form
of a $2\times 2$ matrix equation,
\begin{equation}
[\calH,\calR] = 0\,.
\label{hfb}
\end{equation}

What is relevant for our purpose is the spectrum of the lowest lying
quasiparticles, which for a sufficiently small gap can be obtained
within the BCS approximation, which is much simpler than the solution
of the full HFB equation (\ref{hfb}). We choose a basis in which $h$
is diagonal, i.e., $h_{n\np} = h_n\delta_{n\np}$. Then, within the BCS
approximation, $\rho$ and $\kappa$ are diagonal, too, and given by
\begin{gather}
\rho_n = \frac{1}{2}-\frac{h_n}{2E_n}[1-2f(E_n)]\,,\label{rhobcs}\\
\kappa_n = \frac{\Delta_n}{2E_n}[1-2f(E_n)]\,.\label{kappabcs}
\end{gather}
The quasiparticle energies $E_n = \sqrt{h_n^2+\Delta_n^2}$ and the
quasiparticle occupation numbers $f(E_n) = 1/[\exp(E_n/T)+1]$ are
determined by the diagonal matrix elements $\Delta_n\equiv
\Delta_{nn}$ alone. If we neglect the non-diagonal matrix elements of
$\Delta$, which are irrelevant for the excitation spectrum and, apart
from that, much smaller than the diagonal ones, we can rewrite
\Eqs{rhobcs} and (\ref{kappabcs}) in the compact form
\begin{equation}
\calR = \frac{1}{2}
  -\frac{\calH}{2E}[1-2f(E)]\,.
\label{rbcs}
\end{equation}
It is evident that the generalized density matrix $\calR$ given by
\Eq{rbcs} solves the HFB equation (\ref{hfb}) if $h$ and $\Delta$ are
assumed to be diagonal.

For the spherical case ($\omega_{0x} = \omega_{0y} = \omega_{0z}$) and
moderate numbers of particles ($N \lesssim 10^4$), the self-consistent
HFB equation can be solved numerically \cite{BruunCastin}. However,
for the deformed case and large numbers of particles (experimentally
numbers of the order $N \approx 10^5\dots 10^6$ have been reached),
even within the BCS approximation, the self-consistent solution
becomes numerically intractable. Therefore it may be indicated to
apply semiclassical approximations. Semiclassical methods can become
very accurate for large numbers of particles, and in addition they
often allow for a very clear interpretation of the results. To that
end we will use the Wigner transforms of the density matrix $\rho$,
the pairing tensor $\kappa$, the single-particle hamiltonian $h$,
etc. The Wigner transform of a single-particle operator $A$ is defined
as
\begin{equation}
A(\rv,\pv) = \int\! d^3s\, e^{-i\pv\cdot\sv/\hbar}
  A\Big(\rv+\frac{\sv}{2},\rv-\frac{\sv}{2}\Big)\,.
\label{defwigner}
\end{equation}
The Wigner transform $h(\rv,\pv)$ of the single-particle hamiltonian
$h$ is particularly simple: It is just the classical hamiltonian. We
also recall the useful relations $[A^\dagger](\rv,\pv) = A^*(\rv,\pv)$
and $[\bar{A}](\rv,\pv) = A(\rv,-\pv)$. One advantage of the Wigner
transforms in semiclassical calculations is the product rule for the
Wigner transform of the product of two operators $A$ and $B$
\cite{RingSchuck}, directly leading to an $\hbar$ expansion:
\begin{equation}
[AB](\rv,\pv) = A(\rv,\pv)\,
  \exp\!\Big(\frac{i\hbar\poisson}{2}\Big)\,B(\rv,\pv)\,,
\end{equation}
where the symbol $\poisson$ stands for the Poisson bracket
\begin{equation}
\poisson = \sum_{i=xyz}
      \Big(\frac{\overset{\leftarrow}{\partial}}{\partial r_i}
           \frac{\overset{\rightarrow}{\partial}}{\partial p_i}
          -\frac{\overset{\leftarrow}{\partial}}{\partial p_i}
           \frac{\overset{\rightarrow}{\partial}}{\partial r_i}\Big)\,.
\label{productrule}
\end{equation}

From the definition (\ref{defwigner}) it is clear that the local
density can be written as
\begin{equation}
\rho(\rv) = \rho(\rv,\rv) = \int\! \frac{d^3 p}{(2\pi\hbar)^3}\,
  \rho(\rv,\pv)\,,\label{rholocal}
\end{equation}
As a very simple case we consider the Thomas-Fermi ($\hbar\to 0$)
limit for the density matrix without pairing correlations (i.e.,
$\Delta = \kappa = 0$) at zero temperature. Quantum-mechanically the
density matrix is in this case just given by the Fermi sea filled up
to the Fermi energy $\mu$, i.e., $\rho = \theta(-h)$. To leading order
in $\hbar$ the Wigner transform of this expression gives
$\rho(\rv,\pv) = \theta[-h(\rv,\pv)]$. The corresponding (local)
density reads
\begin{equation}
\rho(\rv) = \frac{p_F^3(\rv)}{6\pi^2\hbar^3}\,,
\label{rhotf}
\end{equation}
with the local Fermi momentum
\begin{equation}
p_F(\rv) = \sqrt{2m[\mu-V(\rv)]}\,\theta[\mu-V(\rv)]\,.
\label{pflocal}
\end{equation}
\Eq{rhotf} together with \Eq{vmeanfield} can easily be solved
self-consistently \cite{Houbiers,Farine}. Since the pairing gaps and
temperatures considered in this article are very small compared with
the Fermi energy, we will use \Eq{rhotf} also in the presence of
pairing correlations and at non-zero temperatures. (The effect of
pairing correlations and temperature on the density profile
$\rho(\rv)$ was investigated in \Ref{Houbiers}). Furthermore, in this
article we are not interested in the details of the density profile
$\rho(\rv)$. As shown in \Ref{Farine}, the self-consistent solution
for $\rho(\rv)$ can be described to good accuracy by approximating the
self-consistent potential $V(\rv)$ again by a harmonic potential
\begin{equation}
V(\rv) = \sum_{i=xyz}\frac{m\omega_i^2}{2}r_i^2\,,
\label{vharm}
\end{equation}
with ``effective'' frequencies $\omega_i > \omega_{0i}$ (we consider
an attractive interaction, i.e., $g > 0$) and an appropriately
readjusted chemical potential $\mu$. In the remaining part of this
article we will use the approximate potential (\ref{vharm}).

In order to include the pairing correlations, one can also use the HFB
equation (\ref{hfb}) in the limit $\hbar\to 0$:
\begin{equation}
[\calH(\rv,\pv),\calR(\rv,\pv)] = 0\,.
\end{equation}
This implies that, to leading order in $\hbar$, at each point $\rv$
the solution $\calR(\rv,\pv)$ as a function of $\pv$ is given by the
solution for a homogeneous system with the density corresponding to
the local density at this point $\rv$ [Local-Density Approximation
(LDA)]:
\begin{equation}
\calR(\rv,\pv) = \frac{1}{2}
  -\frac{\calH(\rv,\pv)}{2E(\rv,\pv)}
  \big(1-2f[E(\rv,\pv)]\big)\,,
\label{rlda}
\end{equation}
with the definition $E(\rv,\pv)=\sqrt{h^2(\rv,\pv)+\Delta^2(\rv)}$.
In terms of the Wigner transform $\kappa(\rv,\pv)$, the gap equation
[\Eq{gapeqcoo}] can be written as
\begin{equation}
\Delta(\rv) = g\int\! \frac{d^3 p}{(2\pi\hbar)^3}\,
  \kappa(\rv,\pv)\,.\label{gapeq}
\end{equation}
Inserting the expression for $\kappa(\rv,\pv)$ corresponding to
\Eq{rlda} into \Eq{gapeq}, we obtain the following non-linear equation
for the gap:
\begin{equation}
\Delta(\rv) = g\int\!\frac{d^3p}{(2\pi\hbar)^3}\,
  \frac{\Delta(\rv)}{2E(\rv,\pv)}\big(1-2f[E(\rv,\pv)]\big)\,.
\label{gapeqlda}
\end{equation}
As mentioned before, this equation is divergent and needs some
regularization (see \Refs{Houbiers,Farine,SadeMelo} for details).

Contrary to the Thomas-Fermi approximation for the unpaired density
matrix, \Eq{rhotf}, which is valid if the potential can be regarded as
constant on a length scale of the inverse Fermi momentum, the
local-density approximation in the paired case is valid only if the
potential is also constant on a length scale of the coherence length
of the Cooper pairs. This latter condition is often not
fulfilled. Therefore, in \Refs{Farine,FarineHekking} an alternative
semiclassical method for the calculation of the gap has been proposed,
which, however, results in an average gap (more precisely: gap
averaged over the Fermi surface) of almost the same magnitude as the
average gap obtained within the local density approximation.
%
\section{Linear response to a slow rotation}
%
\label{linres}
In this section we will describe the formalism used for the
calculation of the moment of inertia of a superfluid gas of trapped
fermionic atoms. Looking at a system rotating with angular velocity
$\Omega$ around the $z$ axis, we can calculate the moment of inertia
from
\begin{equation}
\Theta = \frac{\langle L_z\rangle}{\Omega}
  = \frac{2}{\Omega}\int\! \frac{d^3 r\, d^3 p}{(2\pi\hbar)^3}\,
  (r_x p_y-r_y p_x)\, \rho(\rv,\pv)\,,
\end{equation}
where $\rho(\rv,\pv)$ is the density matrix of the rotating
system. Hence the main problem in calculating the moment of inertia is
to calculate $\rho(\rv,\pv)$, from which also other interesting
quantities like the current density (per spin state)
\begin{equation}
\jv(\rv) = \int\! \frac{d^3 p}{(2\pi\hbar)^3}\,
  \frac{\pv}{m}\rho(\rv,\pv)
\label{defcurrent}
\end{equation}
and the velocity field $\vv(\rv) = \jv(\rv)/\rho(\rv)$
can be derived.

The system is put into rotation by rotating the external trapping
potential around the $z$ axis (of course, for this purpose the
trapping potential must not be axially symmetric). In the rotating
frame, however, the system is still in a stationary state. In this
frame, the Hamiltonian receives the additional term
\begin{equation}
h_1 = -\bar{h}_1 = -\Omega L_z\,,
\end{equation}
which for sufficiently small $\Omega$ can be treated as a
perturbation. This perturbation induces a change of the density
matrix, $\rho_1$, and of the pairing tensor, $\kappa_1$. The mean
field potential is not changed to linear order in $\Omega$, since
$L_z$ is a time-odd operator. Linearizing \Eq{hfb}, we obtain
\begin{equation}
[\calH,\calR_1] = -[\calH_1,\calR]\,,
\label{lhfb}
\end{equation}
where $\calH$ and $\calR$ denote the unperturbed quantities, while
$\calH_1$ and $\calR_1$ refer to the deviations. Assuming that the
unperturbed quantities $\rho$, $\kappa$, $h$, and $\Delta$ are
diagonal (BCS approximation), we can solve \Eq{lhfb} for $\rho_1$ and
$\kappa_1$. (This is equivalent to solving the linearized Gorkov
equations for the normal and anomalous Green's functions at equal
times; see, e.g., \Ref{Valatin}.) The solution reads:
\begin{gather}
\rho_{1\,n\np} = F^{\rho h}_{n\np} h_{1\,n\np}
  + F^{\rho \Delta}_{n\np} \Delta_{1\,n\np}\,,
\label{rho1}\\
\kappa_{1\,n\np} = F^{\kappa h}_{n\np} h_{1\,n\np}
  + F^{\kappa \Delta}_{n\np} \Delta_{1\,n\np}\,,
\label{kappa1}
\end{gather}
where (with the short-hand notation $\rho = \rho_n$, $\rho^\prime =
\rho_\np$, $h = h_n$, $h^\prime = h_\np$, $\kappa = \kappa_n$,
$\kappa^\prime = \kappa_\np$, etc.)
\begin{gather}
F^{\rho h}_{n\np} =
  \frac{(\rho-\rho^\prime)(h+h^\prime)
  -(\kappa-\kappa^\prime)(\Delta+\Delta^\prime)}{E^2-E^{\prime 2}}\,,
\label{frhdef}\\
F^{\rho\Delta}_{n\np} =
  \frac{(\rho+\rho^\prime-1)(\Delta+\Delta^\prime)
  +(\kappa+\kappa^\prime)(h+h^\prime)}{E^2-E^{\prime 2}}\,,
\label{frddef}\\
F^{\kappa h}_{n\np} =
  \frac{(\rho-\rho^\prime)(\Delta-\Delta^\prime)
  +(\kappa-\kappa^\prime)(h-h^\prime)}{E^2-E^{\prime 2}}\,,
\label{fkhdef}\\
F^{\kappa\Delta}_{n\np} =
  \frac{(1-\rho-\rho^\prime)(h-h^\prime)
  +(\kappa+\kappa^\prime)(\Delta-\Delta^\prime)}{E^2-E^{\prime 2}}\,,
\label{fkddef}
\end{gather}

In practice, \Eq{kappa1} is an integral equation, since the change of
the gap, $\Delta_1$, on the r.h.s.\@ is related to the change of the
pairing tensor, $\kappa_1$, by the gap equation. In analogy to
\Eq{gapeq} the gap equation for the perturbed quantities reads
\begin{equation}
\Delta_1(\rv) = g \int\! \frac{d^3p}{(2\pi\hbar)^3}\,
\kappa_1(\rv,\pv)\,.
\label{gapeq1}
\end{equation}

The solution of this integral equation contains some subtleties. For
example, the divergence appearing in \Eq{gapeq1} as a consequence of
the zero-range interaction has to be regularized in the same way as
the correponding divergence of the unperturbed gap equation
(\ref{gapeq}) (see appendix), and the derivations of Eq.\@ (4.34) in
\Ref{Farine}, or the second equation after Eq.\@ (16) in \Ref{Migdal}
are not very explicit about this point. However, these problems can be
circumvented in the following way \cite{Betbeder,Woelfle,Gulminelli}:
Suppose all single-particle wave functions are multiplied by the same
local phase $\exp[i\phi(\rv)]$. Then the HFB equation (\ref{hfb}) can
be rewritten in terms of the gauge-transformed matrices
\begin{gather}
\calRtil = e^{i\Phi}\calR e^{-i\Phi}\,,\\
\calHtil = e^{i\Phi}\calH e^{-i\Phi}\,,
\end{gather}
where
\begin{equation}
\Phi =\Big(\begin{array}{cc}\phi&0\\0&-\phi\end{array}\Big)
\end{equation}
We will consider $\phi$ as small, i.e., of the order of the
perturbation. Then, to linear order in the perturbation, the gauge
transformed HFB equation reads
\begin{equation}
[\calH,\calRtil_1] = -[\calHtil_1,\calR]\,,
\label{glhfb}
\end{equation}
where
\begin{gather}
\calRtil_1 = \calR_1+i[\Phi,\calR]\,,\label{r1g}\\
\calHtil_1 = \calH_1+i[\Phi,\calH]\,.
\end{gather}
In the latter expression one has to take into account that $h$ does
not commute with $\phi$. Explicitly, for a hamiltonian $h$ of the
form (\ref{hsp}) and a local gap $\Delta(\rv)$ one obtains
\begin{gather}
\htil_1 = -\tilde{\bar{h}}_1
  = -\Omega L_z-\frac{\hbar}{2m}\big(\pv\cdot[\nablav\phi(\rv)]
    +[\nablav\phi(\rv)]\cdot\pv\big)\,,
\label{hg}\\
\Deltatil_1(\rv) = \Delta_1(\rv)+2i\phi(\rv)\Delta(\rv)\,.
\label{Deltag}
\end{gather}
Together with the gauge-transformed gap equation
\begin{equation}
\Deltatil_1(\rv) = g \int\! \frac{d^3p}{(2\pi\hbar)^3}\,
  \kappatil_1(\rv,\pv)\,,
\label{ggapeq1}
\end{equation}
\Eq{glhfb} is again a system of integral equations which for an
arbitrary function $\phi(\rv)$ is completely equivalent to the
original one, \Eqs{lhfb} and (\ref{gapeq1}). However, since the
perturbation $\htil_1$ is time-odd, the change of the gap,
$\Deltatil_1$, is purely imaginary and therefore can be eliminated by
an appropriately chosen function $\phi(\rv)$. Physically, this choice
of $\phi(\rv)$ corresponds to a transformation into the local rest
frame of the Cooper pairs \cite{Thouless}. In this particular gauge
the linearized HFB equation reduces to
\begin{gather}
\rhotil_{1\,n\np} = F^{\rho h}_{n\np} \htil_{1\,n\np}\,,
\label{grho1}\\
\kappatil_{1\,n\np} = F^{\kappa h}_{n\np} \htil_{1\,n\np}\,,
\label{gkappa1}
\end{gather}
and instead of \Eq{ggapeq1} we have an equation which determines the
phase $\phi(\rv)$:
\begin{equation}
0\overset{!}{=}g \int\! \frac{d^3p}{(2\pi\hbar)^3}\,
  \kappatil_1(\rv,\pv)\,.
\label{gphieq}
\end{equation}

We now proceed to the evaluation of \Eq{grho1}. The unperturbed
quantities $\rho$ and $\kappa$ entering in $F^{\rho h}_{n\np}$
[\Eq{frhdef}] can be rewritten in terms of $h$ and $\kappa$ according
to the BCS relations (\ref{rhobcs}) and (\ref{kappabcs}). In addition,
as in \Ref{Farine}, we replace $\Delta_n$ by its average value at the
Fermi surface, $\Delta$, because $F^{\rho h}$ and all other relevant
quantities are strongly peaked at $\eps_F$. This allows us to write
$F^{\rho h}_{n\np}$ as a function of two energies $\xi=h_n$ and
$\xip=h_\np$:
\begin{multline}
F^{\rho h}(\xi,\xip)
  = \frac{[1-f(E)-f(\Ep)](\Delta^2+\xi\xip-E\Ep)}{2E\Ep(E+\Ep)}\\
    +\frac{[f(E)-f(\Ep)](\Delta^2+\xi\xip+E\Ep)}{2E\Ep(E-\Ep)}\,,
\label{frhxixi1}
\end{multline}
where we have introduced the abbreviations $E=\sqrt{\xi^2+\Delta^2}$
and $\Ep=\sqrt{\xi^{\prime 2}+\Delta^2}$. In contrast to \Ref{Farine},
we will not drop the thermal quasiparticle occupation numbers $f(E)$
and $f(\Ep)$. As described in detail in \Ref{Farine}, the Wigner
transform of an expression like \Eq{grho1} can be evaluated
semiclassically in the following way. First we rewrite \Eq{grho1} as
an operator equation:
\begin{equation}
\rhotil_1 = \int\! d\xi\, d\xip\, F^{\rho h}(\xi,\xip)
  \delta(h-\xi)\htil_1\delta(h-\xip)\,.
\end{equation}
Then we use the Fourier representation for the $\delta$ functions,
i.e., $\delta(h-\xi) = \int\! dt/(2\pi\hbar)\, \exp[(h-\xi)t/\hbar]$,
and obtain
\begin{multline}
\rhotil_1 = \int\! \frac{d\xib\, d\eps\, dT\,dt}{(2\pi\hbar)^2}\,
  F^{\rho h}\Big(\xib+\frac{\eps}{2},\xib-\frac{\eps}{2}\Big)
    e^{-i\xib T/\hbar} e^{-i\eps t/\hbar}\\
      \times e^{ihT/2\hbar}\htil_1(t) e^{ihT/2\hbar}\,,
\label{aboperator}
\end{multline}
where we have introduced the notation
\begin{equation}
\htil_1(t) = e^{iht/\hbar} \htil_1 e^{-iht/\hbar}\,.
\label{aoft}
\end{equation}
To leading order in $\hbar$ the Wigner transform of the product of the
three operators in the second line of \Eq{aboperator} can be expressed
as the product of their Wigner transforms [see \Eq{productrule}]. Then
the integral over $T$ gives a $\delta$ function of the form
$\delta[h(\rv,\pv)-\xib]$ and the integral over $\xib$ becomes
trivial.

However, for the operator product in $\htil_1(t)$ [\Eq{aoft}] we will
not use the product rule. In this sense we resum certain $\hbar$
corrections to all orders. One can also say that, since the Wigner
transform of \Eq{aoft} involves the classical trajectories (see
below), the long-time information is preserved. On the other hand,
developing the Wigner transform of \Eq{aoft} with the product rule
(\ref{productrule}) into powers of $\hbar$ would lead to the
Wigner-Kirkwood $\hbar$ expansion, which is only valid in the
short-time limit (see \Ref{RingSchuck}). The different treatment of
the operator products in \Eqs{aboperator} and (\ref{aoft}) is
necessary for the following reason: The operator $\htil_1$ connects
states with an energy difference of the order $\hbar\omega$. This is
small compared with the Fermi energy, which is the relevant scale for
the variable $\xib$ [since the result $\rhotil_1(\rv,\pv)$ will be
used in integrals over $\pv$], but not necessarily small compared with
the gap $\Delta$, which is the relevant scale for the variable $\eps$
[this point will become clearer when we investigate the function
$F^{\rho h}(\xib+\eps/2,\xib-\eps/2)$ explicitly].

In the case of the effective harmonic oscillator potential
(\ref{vharm}) the Wigner transform of \Eq{aoft} can be calculated
exactly. The result reads
\begin{equation}
[\htil_1(t)](\rv,\pv) = 
\htil_1[\rv^\mathit{cl}(\rv,\pv;t),
      \pv^\mathit{cl}(\rv,\pv;t)]\,,
\end{equation}
where $\rv^\mathit{cl}(\rv,\pv;t)$ and $\pv^\mathit{cl}(\rv,\pv;t)$
are the classical orbits in the potential (\ref{vharm}) corresponding
to the initial conditions $\rv^\mathit{cl}(\rv,\pv;0) = \rv$ and
$\pv^\mathit{cl}(\rv,\pv;0) = \pv$, which are given by
\begin{gather}
r_i^\mathit{cl}(\rv,\pv;t)
 = r_i \cos(\omega_i t)+\frac{p_i}{m\omega_i} \sin(\omega_i t)\,,
\label{rclt}\\
p_i^\mathit{cl}(\rv,\pv;t)
 = p_i \cos(\omega_i t)-m\omega_i r_i \sin(\omega_i t)\,.
\label{pclt}
\end{gather}
Putting everything together, we obtain
\begin{multline}
\rhotil_1(\rv,\pv) = \int\! d\eps\, F^{\rho h}\Big(h(\rv,\pv)
  +\frac{\eps}{2},h(\rv,\pv)-\frac{\eps}{2}\Big)\\
  \times \int\! \frac{dt}{2\pi\hbar}\, e^{-i\eps t/\hbar}
    \htil_1[\rv^\mathit{cl}(\rv,\pv;t),
      \pv^\mathit{cl}(\rv,\pv;t)]\,.
\label{rho1sc}
\end{multline}

Now we proceed to the calculation of the response of $\rhotil_1$ to
the external perturbation $h_1$, neglecting for the moment the
reaction of the pairing field to the rotation, i.e., the
$\pv\cdot\nablav\phi$ terms in \Eq{hg}. In \Ref{Farine} this
contribution was called the ``Inglis-Belyaev term''
$\rho_1^{\mathit{IB}}$. In this case the Fourier transform in the
second line of \Eq{rho1sc} [with $\htil_1$ replaced by $h_1 = -\Omega
(r_x p_y-r_y p_x)$] can easily be evaluated with the aid of \Eqs{rclt}
and (\ref{pclt}). Inserting the result into \Eq{rho1sc} and observing
that $F(\xi,\xip)$ is symmetric under the exchange of its arguments we
obtain [the arguments of $h(\rv,\pv)$ will be suppressed for brevity]
\begin{multline}
\rho_1^{\mathit{IB}}(\rv,\pv) =\\
-\frac{\Omega\omega_-}{2}
  \Big(\frac{r_x p_y}{\omega_y}+\frac{r_y p_x}{\omega_x}\Big)
    F^{\rho h}\Big(h+\frac{\hbar \omega_+}{2},
      h-\frac{\hbar \omega_+}{2}\Big)\\
-\frac{\Omega\omega_+}{2}
  \Big(\frac{r_x p_y}{\omega_y}-\frac{r_y p_x}{\omega_x}\Big)
    F^{\rho h}\Big(h+\frac{\hbar \omega_-}{2},
      h-\frac{\hbar \omega_-}{2}\Big)\,,
\label{rho1ib}
\end{multline}
with the definition
\begin{equation}
\omega_\pm = \omega_y\pm\omega_x\,.
\end{equation}
To simplify the expression (\ref{rho1ib}) further we note that the
distribution function $\rho(\rv,\pv)$ is changed only in the vicinity
of the Fermi surface, provided the Fermi energy is large compared with
$\hbar \omega_\pm$, $\Delta$, and $T$. Formally this can be inferred
from the fact that $F^{\rho h}(\xib+\eps/2,\xib-\eps/2)$ as a function
of $\xib$ is strongly peaked at $\xib = 0$, which leads us to the
approximation
\begin{equation}
F^{\rho h}(\xib+\eps/2,\xib-\eps/2)
  \approx\Big[G\Big(\frac{\eps}{2\Delta}\Big)-1\Big]\delta(\xib)\,,
\label{appfermi}
\end{equation}
with
\begin{equation}
G(x) = 1+\int\! d\xib\, F^{\rho h}(\xib+x\Delta,\xib-x\Delta)\,.
\label{def1gx}
\end{equation}
At zero temperature the integral in \Eq{def1gx} can be evaluated
analytically, whereas the terms containing the quasiparticle
occupation numbers $f(E)$ and $f(\Ep)$ have to be integrated
numerically. After some manipulations the function $G(x)$ can be
written as
\begin{equation}
G(x) = \frac{\arsinh(x)}{x\sqrt{1+x^2}}
  +\frac{\Delta}{x}\int_0^\infty\! \frac{d\xib}{\xib}
    \Big(\frac{f(E_+)}{E_+}-\frac{f(E_-)}{E_-}\Big) \,,
\label{exprgx}
\end{equation}
with $E_\pm = \sqrt{(\xib\pm x \Delta)^2+\Delta^2}$. Within the
approximation (\ref{appfermi}) the change of the density matrix
corresponding to the Inglis-Belyaev term finally takes the compact
form
\begin{multline}
\rho_1^{\mathit{IB}}(\rv,\pv) = \Omega\delta[h(\rv,\pv)]\Big[
  r_x p_y\Big(1 - \frac{\omega_+ G_- + \omega_- G_+}
    {\omega_+ + \omega_-}\Big)\\
  -r_y p_x\Big(1 - \frac{\omega_+ G_- - \omega_- G_+}
    {\omega_+ - \omega_-}\Big)\Big]\,,
\label{rho1ibfinal}
\end{multline}
with
\begin{equation}
G_\pm = G\Big(\frac{\hbar \omega_\pm}{2\Delta}\Big)\,.
\label{gpm}
\end{equation}

Now we will consider also the change of the pairing field $\Delta$,
i.e., the phase $\phi(\rv)$. As mentioned before, this phase will be
determined by \Eq{gphieq}, where $\kappatil_1(\rv,\pv)$ is obtained
from the Wigner transform of \Eq{gkappa1}. Again we replace $\Delta_n$
and $\Delta_\np$ entering in $F^{\kappa h}_{n\np}$ by the average
value $\Delta$, which allows us to express $F^{\kappa h}_{n\np}$ as a
function of two energies:
\begin{multline}
F^{\kappa h}(\xi,\xip)
  = -\frac{[1-f(E)-f(\Ep)]\Delta(\xi-\xip)}{2E\Ep(E+\Ep)}\\
    -\frac{[f(E)-f(\Ep)]\Delta(\xi-\xip)}{2E\Ep(E-\Ep)}\,.
\label{fkhxixi1}
\end{multline}
Then the Wigner transform of \Eq{gkappa1} can be calculated
semiclassically as given by \Eq{rho1sc} with $\rhotil_1$ and $F^{\rho
h}$ replaced by $\kappatil_1$ and $F^{\kappa h}$, respectively. As it
was the case for $F^{\rho h}(\xib+\eps/2,\xib-\eps/2)$, the function
$F^{\kappa h}(\xib+\eps/2,\xib-\eps/2)$ is strongly peaked at $\xib =
0$, and we approximate it by
\begin{equation}
F^{\kappa h}(\xib+\eps/2,\xib-\eps/2)
  \approx-\frac{\eps}{2\Delta}
    G\Big(\frac{\eps}{2\Delta}\Big)\delta(\xib)\,,
\end{equation}
with
\begin{equation}
G(x) = -\frac{1}{x}\int\! d\xib\, F^{\kappa h}(\xib+x\Delta,\xib-x\Delta)\,.
\label{def2gx}
\end{equation}
It turns out that the definitions (\ref{def1gx}) and (\ref{def2gx})
indeed define the same function $G(x)$, which is explicitly given by
\Eq{exprgx}. Inserting the Wigner transform of \Eq{gkappa1} into
\Eq{gphieq}, we obtain
\begin{multline}
0 = -g \int\! d\eps\, \frac{\eps}{2\Delta}
  G\Big(\frac{\eps}{2\Delta}\Big)
  \int\! \frac{d^3p}{(2\pi\hbar)^3}\, \delta[h(\rv,\pv)] \\
    \times \int\!\frac{dt}{2\pi\hbar}\, e^{-i\eps t/\hbar}
      \htil_1[\rv^\mathit{cl}(\rv,\pv;t),
      \pv^\mathit{cl}(\rv,\pv;t)]\,.
\label{kappa1sc}
\end{multline}
To solve this equation for the phase $\phi(\rv)$ we make the ansatz
\cite{Farine}
\begin{equation}
\phi(\rv) = \alpha \frac{m r_x r_y}{\hbar}\,.
\label{phiansatz}
\end{equation}
Then the second line of \Eq{kappa1sc} is just the Fourier transform of
$\htil_1 = -\Omega (r_x p_y - r_y p_x) - \alpha (r_x p_y + r_y p_x)$,
which is readily evaluated with the aid of \Eqs{rclt} and
(\ref{pclt}). Due to the $\delta$ functions the remaining integrals
are trivial, and \Eq{kappa1sc} finally becomes
\begin{multline}
0 = \frac{i g m^2 p_F(\rv) r_x r_y}{8\pi^2 \hbar^2 \Delta}\\
  \times [\Omega \omega_+ \omega_- (G_+ + G_-) 
    + \alpha (\omega_+^2 G_+ + \omega_-^2 G_-)]\,,
\label{delta1tilde}
\end{multline}
which has the solution
\begin{equation}
\alpha = -\Omega
  \frac{\omega_+ \omega_- (G_+ + G_-)}{\omega_+^2 G_+ + \omega_-^2 G_-}\,.
\label{alphasol}
\end{equation}
Using this expression we can also calculate the change of the original
pairing field, $\Delta_1$: Since the change of the gauge-transformed
pairing field [\Eq{Deltag}] is zero, the original pairing field is
modified according to
\begin{equation}
\Delta_1(\rv) = -2i\Delta\phi(\rv)
  = -\frac{2i\Delta\alpha m r_x r_y}{\hbar}.
\label{delta1ansatz}
\end{equation}

Having calculated the phase $\phi(\rv)$, we can now evaluate
\Eq{rho1sc} with the full $\htil_1$, i.e., including in addition to
the Inglis-Belyaev term [\Eq{rho1ib}] also the response of the density
matrix to the $\pv\cdot\nablav\phi$ terms. This second contribution to
$\rhotil_1$, which we will call $\rho_1^{M_1}$, is obtained in the
same way as discussed above for the first one, and the result reads
\begin{multline}
\rho_1^{M_1}(\rv,\pv) = \alpha\delta[h(\rv,\pv)]\Big[
  r_x p_y \Big(1-\frac{\omega_+ G_+ + \omega_- G_-}
    {\omega_+ + \omega_-}\Big)\\
 +r_y p_x \Big(1-\frac{\omega_+ G_+ - \omega_- G_-}
    {\omega_+ - \omega_-}\Big)\Big]\,.
\end{multline}
However, we are not interested in the change of the gauge-transformed
density matrix, $\rhotil_1$, but of the original density matrix,
$\rho_1$. According to \Eq{r1g} the relation between $\rho_1$ and
$\rhotil_1$ is given by
\begin{equation}
\rho_1 = \rhotil_1-i[\phi,\rho]
  = \rho_1^\mathit{IB}+\rho_1^{M_1}+\rho_1^{M_2}\,.
\end{equation}
Due to the simple $\rv$ dependence of $\phi$, the Wigner transform of
the commutator $[\phi,\rho]$ is identical to the Poisson bracket of
the Wigner transforms of $\phi$ and $\rho$, i.e.
\begin{align}
\rho_1^{M_2}(\rv,\pv) &= \hbar\phi(\rv)\poisson\rho(\rv,\pv)
  \nonumber\\
  &= \alpha m\Big(r_x \frac{\partial}{\partial p_y} +
        r_y \frac{\partial}{\partial p_x}\Big)\rho(\rv,\pv)\,.
\end{align}
As we did before, we will assume that $\Delta$ and $T$ are much
smaller than the Fermi energy. Therefore we can write $\rho(\rv,\pv)
\approx \theta[-h(\rv,\pv)]$ and we obtain
\begin{equation}
\rho_1^{M_2}(\rv,\pv) =
   -\alpha (r_x p_y + r_y p_x) \delta[h(\rv,\pv)]\,.
\end{equation}
The total effect of the phase $\phi$, i.e., of the reaction of the
pairing field, on the density matrix, which in \Ref{Farine} was called
the ``Migdal term'' $\rho_1^M$, is the sum of the two contributions
$\rho_1^{M_1}$ and $\rho_1^{M_2}$:
\begin{multline}
\rho_1^M(\rv,\pv) = -\alpha\delta[h(\rv,\pv)]\Big(
  r_x p_y \frac{\omega_+ G_+ + \omega_- G_-}
    {\omega_+ + \omega_-}\\
 +r_y p_x \frac{\omega_+ G_+ - \omega_- G_-}
    {\omega_+ - \omega_-}\Big)\,.
\label{rho1mfinal}
\end{multline}
Together with the Inglis-Belyaev term, \Eq{rho1ibfinal}, and the
explicit expression for $\alpha$, \Eq{alphasol}, our final result for
the change of the density matrix reads
\begin{multline}
\rho_1(\rv,\pv) = \Omega \delta[h(\rv,\pv)] \Big(r_x p_y-r_y p_x\\
  -\frac{4 G_+ G_-(\omega_x^2 r_x p_y-\omega_y^2 r_y p_x)}
    {\omega_+^2 G_++\omega_-^2 G_-}\Big)\,.
\label{rho1final}
\end{multline}

Given the change of the density matrix, we can immediately calculate
the current density $\jv(\rv)$ [\Eq{defcurrent}] and the velocity
field $\vv(\rv)$:
\begin{multline}
\jv(\rv) = \rho(\rv)\vv(\rv) = \Omega \rho(\rv)\Big(r_x \ev_y-r_y \ev_x\\
  -\frac{4 G_+ G_-(\omega_x^2 r_x \ev_y-\omega_y^2 r_y \ev_x)}
    {\omega_+^2 G_++\omega_-^2 G_-}\Big)\,.
\label{current}
\end{multline}
It is interesting to check explicitly that this current fulfils the
continuity equation. In the rotating frame the continuity equation
reads
\begin{equation}
\nablav\cdot\jv(\rv) + \dot{\rho}(\rv)
  - \Omega(\ev_z\times\rv)\cdot\nablav\rho(\rv) = 0\,,
\label{continuity}
\end{equation}
where $\dot{\rho}(\rv)=0$ in our case of a stationary rotation. Taking
the divergence of \Eq{current}, we get from the second line a
contribution proportional to $[\nablav \rho(\rv)] \cdot [\ev_z \times
\nablav V(\rv)]$. This is zero, since the gradient of the density in
Thomas-Fermi approximation [\Eq{rhotf}], $\nablav\rho(\rv)$, is
parallel to $\nablav V(\rv)$. Thus, the divergence of the current is
equal to the divergence of the first line of \Eq{current}, which
exactly fulfils \Eq{continuity}. Note that the contribution of the
Migdal term is crucial in order to satisfy the continuity
equation. The easiest way to see this is to consider the limit
$\Delta\to \infty$. In this limit we have $G_\pm \to 1$ and
$\rho_1^\mathit{IB}(\rv,\pv) \to 0$, which implies
$\jv^\mathit{IB}(\rv)\to 0$. Hence, with the Inglis-Belyaev
contribution alone, \Eq{continuity} cannot be satisfied.

As observed in \Ref{Farine}, the velocity field $\vv(\rv)$ describes a
mixture of rotational motion, corresponding to a velocity field
proportional to $\ev_z\times \rv$, and irrotational motion,
corresponding to a velocity field proportional to $\nablav(r_x
r_y)$. The ordinary rigid rotation is realized if $G_+ = G_- =
0$. This is the case if the temperature approaches the critical
temperature $T_c$, where the gap vanishes [the temperature dependence
of the function $G(x)$ will be discussed in \Sec{results}], but it can
also happen at zero temperature if $\Delta\ll \hbar\omega_\pm$, as
discussed in \Ref{Farine}. Purely irrotational motion, as it is
expected in homogeneous superfluids, is reached if $G_+ = G_- =
1$. This is only possible if the temperature is very low and if
$\Delta\gg \hbar\omega_\pm$.

For completeness let us also discuss the change of the pairing tensor,
$\kappa_1(\rv,\pv)$, which can be obtained in a way completely
analogous to the calculation of the change of the density matrix,
$\rho_1(\rv,\pv)$. The result reads
\begin{multline}
\kappa_1(\rv,\pv) = \frac{2i\hbar\Omega}{m\Delta}\,
  \frac{\omega_+\omega_-G_+G_-}{\omega_+^2G_++\omega_-^2G_-} 
    p_xp_y\delta[h(\rv,\pv)]\\
  -2i\phi(\rv)\,\kappa(\rv,\pv)\,.
\end{multline}
Since the last term is of the order $\hbar^{-1}$ [see \Eq{phiansatz}],
it has been argued that a semiclassical description is possible only
in the particular gauge where $\Delta+\Delta_1$ is real and where this
term vanishes \cite{Woelfle,Gulminelli}.
%
\section{Superfluid rotation in transport theory}
%
\label{transport}
The transport or hydrodynamical equations for superfluid systems can
be derived by taking the $\hbar\to 0$ limit of the time-dependent
Hartree-Fock-Bogoliubov (TDHFB) equation
\cite{Woelfle,RingSchuck,Valatin}
\begin{equation}
i\hbar \dot{\calR} = [\calH,\calR]\,,
\label{tdhfb}
\end{equation}
i.e., by replacing the Wigner transforms of the commutators in
\Eq{tdhfb} by Poisson brackets of the Wigner transforms
\cite{Betbeder,Serene,DiToro,Woelfle,Gulminelli}. Due to the
transformation into the rotating frame, we are dealing with a static
problem, where the TDHFB equation (\ref{tdhfb}) reduces to the HFB
equation (\ref{hfb}). Again we make use of the gauge transformation
and retain only terms of linear order in the perturbation. Then, if
$\phi(\rv)$ is chosen such that $\Deltatil_1$ vanishes [\Eq{gphieq}],
the leading order in $\hbar$ of \Eq{glhfb} becomes
\begin{gather}
i\hbar h\poisson \rhotil_1 + 2\Delta \kappatil_1
  = -i\hbar \htil_1\poisson \rho\,,
\label{trans1}\\
i\hbar \Delta\poisson \rhotil_1 - 2h \kappatil_1
  = i\hbar \htil_1\poisson \kappa\,.
\label{trans2}
\end{gather}
In this equation and in the remaining part of this section, $h$,
$\rho$, $\kappa$, etc.\@ denote the Wigner transforms of the
corresponding operators; the arguments $\rv$ and $\pv$ are suppressed
for brevity.

Let us first study the zero-temperature limit, $T = 0$. In this case
the unperturbed quantities are given by $\rho = (1-h/E)/2$ and $\kappa
= \Delta/2E$ [see \Eq{rlda} in the limit $T\to 0$], and it is easy
to show that $(\htil_1\poisson\rho)h =
(\htil\poisson\kappa)\Delta$. Thus, for $\Delta\not= 0$, the solution
of \Eqs{trans1} and (\ref{trans2}) reads
\begin{gather}
\rhotil_1 = 0\,,\\
\kappatil_1 = -\frac{i\hbar}{2\Delta}(\htil_1\poisson\rho)\,.
\label{kappa1trans}
\end{gather}
As we will see, the relation $\rhotil_1 = 0$ implies that the velocity
field is completely irrotational independent of the magnitude of
$\Delta$, which is a well-known property of homogeneous superfluid
systems at $T = 0$.

Now we are going to determine the phase $\phi$. To that end we insert
\Eq{kappa1trans} into \Eq{gphieq}. If we make again the ansatz
(\ref{phiansatz}), we obtain the following equation:
\begin{equation}
0 = -\frac{ig\hbar}{2\Delta}\int\! \frac{d^3p}{(2\pi\hbar)^3}
  \Big((\Omega+\alpha)r_x \frac{\partial\rho}{\partial r_y}
      -(\Omega-\alpha)r_y \frac{\partial\rho}{\partial r_x}\Big)\,.
\label{phitrans}
\end{equation}
[Note that in this equation $\rho$ still refers to the Wigner
transform of the non-local density matrix, $\rho(\rv,\pv)$.] It is
clear that in general this equation does not have a solution for all
$\rv$, since the ansatz (\ref{phiansatz}) is not general enough. But
under certain assumptions it turns out that this ansatz is
sufficient. Firstly, we assume that the gap $\Delta(\rv)$ is either
replaced by a constant corresponding to its average value at the Fermi
surface (as it was done in the previous section), or that
$\Delta(\rv)$ is calculated within the LDA. In these both cases the
function $\Delta(\rv)$ can formally be written as
$\Delta[V(\rv)]$. Using this, we define the following short-hand
notation:
\begin{equation}
\frac{d\rho}{dV} =
  \frac{d\rho}{dh}\,\frac{dh}{dV}+\frac{d\rho}{d\Delta}\,\frac{d\Delta}{dV}
  = -\frac{\Delta^2}{2E^3}+\frac{h\Delta}{2E^3}\,\frac{d\Delta}{dV}\,,
\end{equation}
which allows us to write $\nablav\rho = (d\rho/dV) \nablav V$.
Secondly, as in the previous section, we assume that the potential
$V(\rv)$ is a harmonic oscillator. Then \Eq{phitrans} becomes
\begin{multline}
0 = -\frac{ig\hbar mr_x r_y}{2\Delta}
  \int\! \frac{d^3p}{(2\pi\hbar)^3}\,\frac{d\rho}{dV}\\
  \times [\Omega(\omega_y^2-\omega_x^2)+\alpha(\omega_y^2+\omega_x^2)]\,,
\end{multline}
with the solution
\begin{equation}
\alpha = \alpha_0
       = -\Omega \frac{\omega_y^2-\omega_x^2}{\omega_y^2+\omega_x^2}\,.
\label{alpha0}
\end{equation}
Not surprisingly, this result is identical to the $\hbar\to 0$ limit
of \Eq{alphasol}, since for $T = 0$ we have $G(0) = 1$ and consequently
$\lim_{\hbar\to 0}G_\pm = 1$.

As in the previous section, the phase $\phi$ implies a change of the
density matrix, $\rho_1$, due to the inverse gauge transformation, which
to leading order in $\hbar$ reads
\begin{equation}
\rho_1 = \rhotil_1 + \hbar \phi\poisson\rho\,.
\label{invgaugetrans}
\end{equation}
As we have seen, the first term vanishes. Thus, to linear order in the
perturbation, \Eq{invgaugetrans} can be rewritten in the following,
more suggestive way:
\begin{equation}
\rho(\rv,\pv)+\rho_1(\rv,\pv) =
  \rho[\rv,\pv+\hbar\nablav\phi(\rv)]\,.
\end{equation}
From this equation it follows immediately that the velocity field is
given by
\begin{equation}
\vv(\rv) = -\frac{\hbar}{m} \nablav\phi(\rv)\,,
\label{virrot}
\end{equation}
which is completely irrotational. Note that this result does not
depend on the form of $\phi(\rv)$ and the approximations made to
calculate $\phi(\rv)$. It also does not at all depend on the magnitude
of $\Delta$, as long as $\Delta\not= 0$. It is rather a direct
consequence of the vanishing of $\rhotil_1$, which in turn follows
immediately from the $\hbar\to 0$ limit of the linearized HFB
equations for time-odd perturbations and zero temperature. However, as
we have seen in the previous section, in a small system where
$\hbar\omega$ is of the same order of magnitude as $\Delta$, the
velocity field is not irrotational. Our conclusion is that one should
be careful when applying transport theory to such systems.

So far we have considered only the zero-temperature limit. In the
remaining part of this section we are going to consider also the case
$T>0$. In this case it is difficult to solve the coupled \Eqs{trans1}
and (\ref{trans2}). However, if we in analogy to the previous section
assume that the the unperturbed gap $\Delta$ is constant, we find the
following solution for $\rhotil_1$ and $\kappatil_1$:
\begin{gather}
\rhotil_1 =
  \Big(\frac{d\rho}{dh}-\frac{\Delta}{h}\frac{d\kappa}{dh}\Big)\htil_1
    = \frac{df(E)}{dE}\htil_1\,,
\label{rhotiltrans}\\
\kappatil_1 = -\frac{i\hbar}{2h}\frac{d\kappa}{dh}(\htil_1\poisson h)\,.
\label{kappatrans}
\end{gather}
If we again make the ansatz (\ref{phiansatz}) and insert
\Eq{kappatrans} into \Eq{gphieq}, we find $\alpha = \alpha_0$ as in
the zero-temperature case [see \Eq{alpha0}]. This could have been
anticipated from the $\hbar\to 0$ limit of \Eq{alphasol}, which does
not depend on the actual value of $G(0)$. Finally we are now going to
calculate $\rho_1$. To that end we insert \Eqs{rhotiltrans} and
(\ref{phiansatz}) with $\alpha = \alpha_0$ into \Eq{invgaugetrans},
and we obtain
\begin{multline}
\rho_1 = -\Omega \frac{df(E)}{dE}(r_x p_y-r_y p_x)\\
  -\alpha_0\Big(\frac{df(E)}{dE}
    -\frac{d\rho}{dh}\Big)(r_x p_y+r_y p_x)\,.
\label{rhotrans}
\end{multline}
Since $df(E)/dE$ and $d\rho/dh$ are both strongly peaked at the Fermi
surface, we can make the same approximation as in the previous
section, i.e., we replace the strongly peaked functions by $\delta$
functions with the appropriate strength. Noting that
\begin{equation}
\lim_{x\to 0} G(x) = 1+\int\! d\xi\, \frac{\Delta^2}{\xi}\,
  \frac{d}{d\xi}\frac{f(E)}{E} = 1+\int\! d\xi\, \frac{df(E)}{dE}\,,
\end{equation}
we can write the result as
\begin{multline}
\rho_1 = \Omega [1-G(0)]\delta(h)(r_x p_y-r_y p_x)\\
-\alpha_0G(0)\delta(h)(r_x p_y+r_y p_x)\,,
\end{multline}
which is in perfect agreement with the $\hbar\to 0$ limit of
\Eqs{rho1ibfinal} and (\ref{rho1mfinal}).
%
\section{Results and discussion}
%
\label{results}
Using the results for change of the non-local density matrix
$\rho_1(\rv,\pv)$ given in the previous sections, we can now calculate
the moment of inertia. It should be remembered that an ideal Fermi gas
at zero temperature behaves like a rigid body, i.e., the velocity
field is given by $\vv(\rv) = \Omega \ev_z \times \rv$. Since the
critical temperature for the BCS transition is very low, $\Theta$ will
approach the rigid-body value $\Theta_\mathit{rigid}$ for $T\to
T_c$. Using the Thomas-Fermi density profile (\ref{rhotf}) with the
effective harmonic oscillator potential (\ref{vharm}), we can
immediately calculate $\Theta_\mathit{rigid}$. The result reads
\begin{equation}
\Theta_\mathit{rigid}=\frac{\mu^4(\omega_x^2+\omega_y^2)}
  {12\hbar^3\omega_x^3\omega_y^3\omega_z}\,.
\end{equation}
In terms of $\Theta_\mathit{rigid}$ the moment of inertia of the
superfluid system as obtained from $\rho_1(\rv,\pv)$ can be written as
\begin{equation}
\Theta = \Theta_\mathit{rigid} \Big(1-\frac{8\omega_x^2\omega_y^2 G_+ G_-}
  {(\omega_x^2+\omega_y^2)(\omega_+^2G_++\omega_-^2G_-)}\Big)\,.
\label{theta}
\end{equation}
In the $\hbar\to 0$ (transport) limit, where $G_\pm\to G(0)$,
\Eq{theta} reduces to
\begin{equation}
\Theta = \Theta_\mathit{rigid}\Big[1-G(0)+G(0)
  \Big(\frac{\omega_y^2-\omega_x^2}{\omega_y^2+\omega_x^2}\Big)^{\!2}\Big]\,.
\label{thetatrans}
\end{equation}

In fact, this formula can be understood very easily. The moment of
inertia corresponding to the purely irrotational velocity field as it
is expected for a large superfluid system at zero temperature,
$\vv(\rv) = -\alpha_0\nablav(r_x r_y)$, is given by
\begin{equation}
\Theta_\mathit{irrot} = \Theta_\mathit{rigid}
  \Big(\frac{\omega_y^2-\omega_x^2}{\omega_y^2+\omega_x^2}\Big)^{\!2}\,.
\end{equation}
Within the two-fluid model a homogeneous system of density $\rho$ is
described as a mixture of a superfluid component of density $\rho_s$
and a normal-fluid component of density $\rho_n$, with $\rho_s+\rho_n
= \rho$. At $T = 0$ one has $\rho_s = \rho$ and $\rho_n = 0$, whereas
at $T \geq T_c$ one has $\rho_s = 0$ and $\rho_n = \rho$. If this
model was correct also for finite systems, one would expect that the
moment of inertia is given by
\begin{equation}
\Theta = \frac{\rho_n}{\rho} \Theta_\mathit{rigid} + 
  \frac{\rho_s}{\rho} \Theta_\mathit{irrot}\,.
\end{equation}
This would be exactly \Eq{thetatrans}, if we could identify $G(0)$
with $\rho_s/\rho$. In fact, the microscopic calculation of
$\rho_s$ for a homogeneous system gives \cite{FetterWalecka}
\begin{equation}
\rho_s = \rho-\frac{1}{6\pi^2m\hbar^3}\int_0^\infty\!
  dp\,p^4\Big(-\frac{df(E)}{dE}\Big)\,,
\end{equation}
with $E = \sqrt{(p^2/2m-\mu)^2+\Delta^2}$. Noting that the integrand
is peaked at $p = p_F$ and remembering $\rho = p_F^3/6\pi^2\hbar^3$,
we rewrite this as
\begin{equation}
\frac{\rho_s}{\rho} \approx 1+\int\! d\xi\,\frac{df(E)}{dE}\,.
\end{equation}
with $\xi = p^2/2m-\mu$. As noted in \Sec{transport}, the r.h.s.\@ of
this equation is identical to $\lim_{x\to0} G(x)$, so that we are left
with
\begin{equation}
\frac{\rho_s}{\rho} = G(0)\,.
\end{equation}

The previous paragraph can be summarized in the statement that the
transport approach, corresponding to the leading order of the $\hbar$
expansion, reproduces the two-fluid model for homogeneous systems. It
does not give any finite-size corrections, as can be seen from the
fact that the result does not depend on the trapping frequencies,
except for the purely geometrical dependence contained in
$\Theta_\mathit{rigid}$ and $\Theta_\mathit{irrot}$. In contrast to
this, the method described in \Sec{linres} is capable to describe the
different behavior of the system depending on whether the trapping
frequencies (multiplied by $\hbar$) are small or large compared with
the gap $\Delta$. This dependence is governed by the $G_\pm$ factors
appearing in \Eq{theta}, resulting from the long-time behavior of the
operator $\htil_1(t)$ [see discussion after \Eq{aoft}]. In order to
reproduce this behavior within the $\hbar$ expansion, one would have
to resum a certain class of corrections proportional to
$\hbar\omega_\pm/\Delta$ to all orders, in particular if one wants to
cover the whole range of possible parameters from $\hbar\omega_\pm
\ll\Delta$ to $\hbar\omega_\pm\gg\Delta$.

\begin{figure}
\epsfig{file=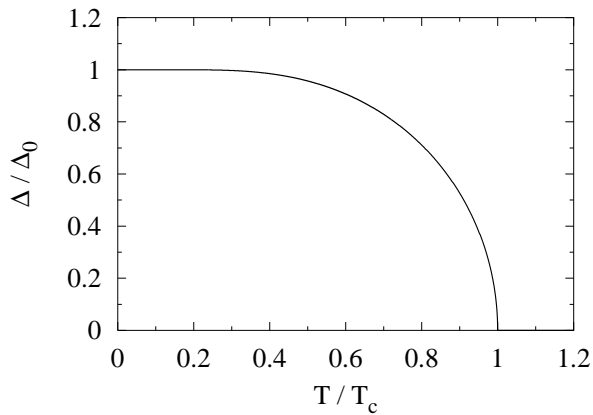,width=8cm}
\caption{Universal function for the temperature dependence of the
gap ($T_c = 0.567 \Delta_0$).\label{figdeltat}}
\end{figure}

Let us now proceed to a quantitative analysis. In order to calculate
the moment of inertia $\Theta$ as a function of temperature, we need
the temperature dependence of the gap $\Delta$. As in \Ref{Farine}, we
will assume that it is described by the same universal function
relating $\Delta/\Delta_0$ to $T/T_c$ in homogeneous matter, where
$\Delta_0$ denotes the gap at $T = 0$ and $T_c = 0.567 \Delta_0$. This
universal function is given by the solution of the non-linear equation
\cite{Lifshitz}
\begin{equation}
-\ln\!\Big(\frac{\Delta}{\Delta_0}\Big)
  = \int\! d\xi\, \frac{f(E)}{E}\,.
\end{equation}
For completeness it is displayed in \Fig{figdeltat}.

\begin{figure}
\epsfig{file=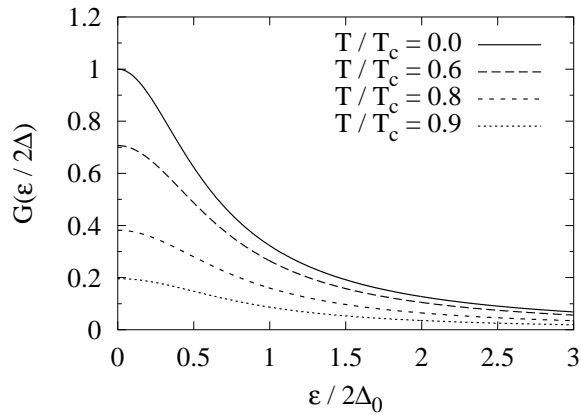,width=8cm}
\caption{Behavior of the function $G(\eps/2\Delta)$ for different
temperatures between $0$ and $T_c$.\label{figfe2d}}
\end{figure}

For the calculation of the moment of inertia we also need the function
$G(\eps/2\Delta)$, which depends on $T$ via the temperature dependence
of $\Delta$ discussed above, and via the explicit temperature
dependence of the function $G(x)$ due to the thermal quasiparticle
occupation numbers as given by \Eq{exprgx}. If only the temperature
dependence of $\Delta$ was included, $G(\eps/2\Delta)$ as a function
of $\eps$ would become very strongly peaked at $\eps = 0$ for $T\to
T_c$. However, due to the explicit temperature dependence of the
function $G(x)$, the peak is suppressed and as a function of $\eps$
the function $G(\eps/2\Delta)$ even becomes more and more flat with
increasing temperature, as shown in \Fig{figfe2d}.
The decrease of $G(0)$ when $T$ approaches $T_c$ reflects the decrease
of the superfluid fraction in the two-fluid model.

Next we have to specify the parameters of the system. For our
comparison we consider, as in \Ref{Farine}, $583000$ $^6$Li atoms
(i.e., $286500$ atoms per spin state) in a harmonic oscillator
potential with average frequency $\hbar\omega =
\hbar(\omega_x\omega_y\omega_z)^{1/3} = 8.21\,\mbox{nK}$. The
corresponding chemical potential is $\mu = 983\,\mbox{nK}$. In order
to simulate the effect of the self-consistent mean-field potential,
the frequency has been chosen slightly higher than the frequency of
the external trapping potential ($\hbar\omega_0 = 6.9\,\mbox{nK}$)
\cite{Farine}.

In the experiments the traps are generally very elongated, i.e. we
have a strong deformation $\sigma = \omega_z/\omega_\perp$, where
$\omega_\perp = \sqrt{\omega_x\omega_y}$ is the average frequencey in
the $xy$ plane. In our examples we choose $\sigma = 1/8$. This results
in a rather high value for the average transverse frequency of
$\hbar\omega_\perp = \hbar\omega/\sigma^{1/3} = 16.42\,\mbox{nK}$. In
order to rotate the system around the $z$ axis, at least a small
deformation in the $xy$ plane is necessary, which we parametrize by
$\delta = \omega_x/\omega_y$. (In practice, the rotating deformation
of the potential can be generated by a laser beam \cite{Madison}.)

The main uncertainty comes from the gap at zero temperature,
$\Delta_0$. Note that the coupling constant $g$ does not appear
explicitly. The moment of inertia depends on the interaction only via
$\Delta$, which can be written as a function of $T$ and
$\Delta_0$. The value of the critical temperature $T_c = 0.567
\Delta_0$ is still under investigation. In addition, the $s$-wave
scattering length $a$ of the atoms, and consequently $g$, $\Delta_0$,
and $T_c$, can be tuned in the experiments by a magnetic field due to
the presence of Feshbach resonances. Therefore we will treat
$\Delta_0$ as a free parameter. As a rough estimate, using the
scattering length $a = -2160 a_0$, where $a_0$ is the Bohr radius, one
obtains that the gap $\Delta_0$ averaged over the Fermi surface is of
the order of magnitude of $15\,\mbox{nK}$ \cite{Farine}, i.e., of the
same order of magnitude as the transverse trapping frequency
$\omega_\perp$.

\begin{figure}
\epsfig{file=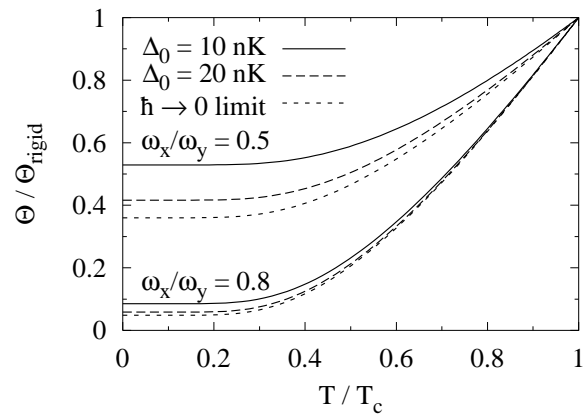,width=8cm}
\caption{Moment of inertia as a function of $T/T_c$ for small
($\omega_x/\omega_y = 0.8$) and large ($\omega_x/\omega_y = 0.5$)
deformations in the $xy$ plane and two values of $\Delta_0$ ($10$ and
$20\,\mbox{nK}$). The short-dashed lines correspond to the $\hbar \to
0$ limit. \label{figtht}}
\end{figure}

In \Fig{figtht}
we display the moment of inertia as a function of the temperature for
two different deformations $\delta$. The lower curves correspond to a
very small deformation, $\delta = 0.8$. In this case the moment of
inertia at $T = 0$ is very small. When $T$ approaches $T_c$, the
normal-fluid component becomes more and more important and
consequently the moment of inertia increases until it finally reaches
the rigid-body value at $T = T_c$. Qualitatively the behavior is
similar in the case of a strong deformation in the $xy$ plane (upper
curves), except that in this case the whole curve is shifted upwards,
mainly due to the much larger value of $\Theta_\mathit{irrot}$. The
difference between the three curves shown for each deformation will be
discussed below.

\begin{figure}
\epsfig{file=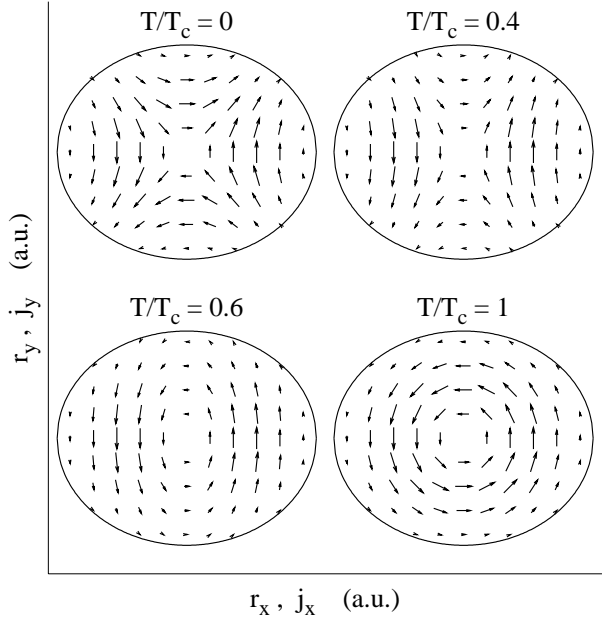,width=8cm}
\caption{Current distributions $\jv(r_x,r_y,0)$ in the $xy$ plane
(arbitrary units) for $\omega_x/\omega_y = 0.8$, $\Delta_0 =
20\,\mbox{nK}$, and four different temperatures: $T/T_c = 0$, $0.4$,
$0.6$, and $1$.
\label{figj}}
\end{figure}

In order to illustrate the origin of the temperature dependence of
$\Theta$, we show in \Fig{figj}
the current distributions for the case $\delta = 0.8$ for four
temperatures between $T = 0$ and $T = T_c$. One can clearly see the
continuous transition from the irrotational motion at $T = 0$,
resulting in a small angular momentum and therefore a small moment of
inertia, to the rigid motion at $T = T_c$.

Now we are going to discuss the differences between the three curves
shown in \Fig{figtht} for each deformation. The short-dashed lines
correspond to the results obtained within the $\hbar\to 0$ approach,
\Eq{thetatrans}. The long-dashed and solid lines were obtained from
\Eq{theta}, i.e., they take into account the difference between
$G_\pm$ and $G(0)$, resulting from the long-time behavior of the
operator $\htil_1(t)$, \Eq{aoft}. From the definition (\ref{gpm}) it is
clear that this difference is less important for large values of
$\Delta$, and indeed the long-dashed lines, corresponding to $\Delta_0
= 20\,\mbox{nK}$, are closer to the $\hbar\to 0$ results than the
solid lines, corresponding to $\Delta_0 = 10\,\mbox{nK}$. More
precisely, the criterion for the validity of the $\hbar\to 0$ approach
seems to be $\hbar \omega_\perp \ll \Delta_0$ rather than $\hbar
\omega_\perp \ll \Delta$, as one might expect. This surprising fact
can be understood by looking at \Fig{figfe2d}: Whatever is the actual
value of the temperature $T$ [i.e., of $\Delta(T)$], the value of
$G(\eps/2\Delta)$ can always be replaced by $G(0)$ if
$\eps/2\Delta_0\ll 1$.

\begin{figure}
\epsfig{file=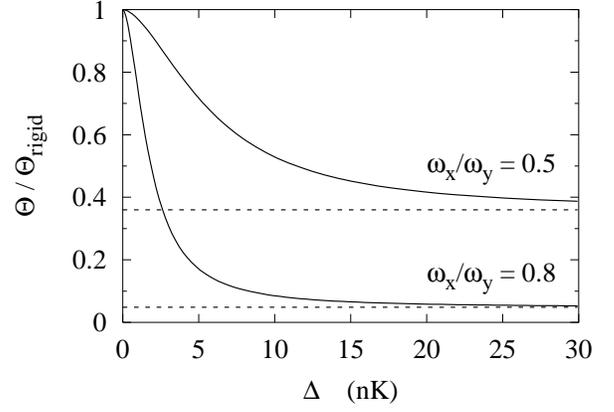,width=8cm}
\caption{Moment of inertia for zero temperature as a function of
$\Delta$ for small ($\omega_x/\omega_y = 0.8$) and large
($\omega_x/\omega_y = 0.5$) deformations in the $xy$ plane. The dashed
lines indicate the corresponding irrotational limits.
\label{figthd}}
\end{figure}

To show more clearly the non-trivial dependence of $\Theta$ on
$\Delta$, we show in \Fig{figthd}
the moment of inertia for zero temperature as a function of $\Delta$
for the same deformations as in \Fig{figtht}. The irrotational limit,
indicated by the dashed lines, is reached for $\Delta\to\infty$. If
$\Delta$ is much smaller than $\hbar \omega_-$ ($3.67\,\mbox{nK}$ in
the case $\delta = 0.8$ and $11.61\,\mbox{nK}$ in the case $\delta =
0.5$, respectively), the moment of inertia even approaches the rigid
body value, and the $\hbar$ expansion fails completely. For example,
in nuclear physics strong deviations from the irrotational value are
quite common \cite{Durand,Migdal}.

Finally let us briefly discuss the question whether the moment of
inertia is suitable to detect the superfluidity in experiments. In
principle the moment of inertia can be measured directly by measuring
the rotational energy
\begin{equation}
E_\mathit{rot} = \frac{\Theta}{2} \Omega^2\,.
\end{equation}
Since the rotation does not change the potential energy (at least not
to linear order in $\Omega$), the rotational energy is equal to the
difference of the release energies $E_\mathit{rel}$ of the rotating
system and of the non-rotating system. (The release energy
$E_\mathit{rel}$ is the total energy of all particles after the
trapping potential has been switched off, i.e., the sum of the kinetic
energy $E_\mathit{kin}$ and of the interaction energy $E_\mathit{int}$
of the trapped system.) A disadvantage of the direct measurement of
$E_\mathit{rot}$ is that it requires two identical systems, one in
rotation and one at rest. As a rough estimate we approximate the
release energy $E_\mathit{rel}$ by the kinetic energy $E_\mathit{kin}$
of the particles in the ``effective'' harmonic potential
(\ref{vharm}),
\begin{equation}
E_\mathit{kin} =
  2\int\! \frac{d^3r\, d^3p}{(2\pi\hbar)^3}\,\frac{\pv^2}{2m} \rho(\rv,\pv)
  = \frac{\mu^4}{8\hbar^3\omega_x\omega_y\omega_z}\,.
\end{equation}
Hence, as a function of the average transverse trapping frequency
$\omega_\perp = \sqrt{\omega_x\omega_y}$ and the deformation $\delta =
\omega_x/\omega_y$ we obtain
\begin{equation}
\frac{E_\mathit{rot}}{E_\mathit{kin}} = 
  \frac{1+\delta^2}{3\delta} \frac{\Theta}{\Theta_\mathit{rigid}}
  \Big(\frac{\Omega}{\omega_\perp}\Big)^2\,.
\end{equation}
Since we used linear response theory, our results are valid only for
slow rotations, $\Omega \ll\omega_\perp$. In particular the angular
velocity must be small enough in order to avoid the creation of
vortices. For an optimistic estimate we choose $\Omega =
0.4\omega_\perp$. Since the difference of $\Theta$ between $T = T_c$
and $T = 0$ is most pronounced for small deformation (see
\Fig{figtht}), we choose $\delta = 0.8$. Using these numbers we find
$E_\mathit{rot} / E_\mathit{kin} \approx 0.1 \times
\Theta / \Theta_\mathit{rigid}$, i.e., the moment of inertia might
indeed be measurable.
%
\section{Summary and conclusions}
%
\label{summary}
In this article we have discussed the temperature dependence of the
moment of inertia of a Fermi gas trapped in a slowly rotating trapping
potential. The assumption of a slow rotation allowed us to use linear
response theory (RPA), but it is clear that in this way certain
interesting effects like the creation of vortices could not be
considered, since they depend non-linearly on the angular velocity
$\Omega$ of the rotation.

In \Sec{linres} we derived the density matrix of the rotating system
using a semiclassical method similar to the one described in
\Ref{Farine}, but now taking into account the thermal quasiparticle
occupation numbers, which were neglected in \Ref{Farine} and which
give rather important contributions. One important point is that the
method takes into account that the energy difference $\hbar
\omega_\pm$ of the states connected by the perturbation hamiltonian
(i.e., essentially by $L_z$) is not necessarily negligible in
comparison with the gap $\Delta$. This leads to a non-trivial behavior
of the density matrix on $\hbar \omega_\pm/\Delta$. These effects can
also be regarded as finite-size effects, since $\hbar \omega_\pm$
vanishes in homogeneous systems.

In \Sec{transport} we presented an alternative method for the
calculation of the density matrix, where only the leading order of the
$\hbar$ expansion is retained. This is equivalent to the transport or
hydrodynamical approach which is often used in the literature
\cite{Zambelli,Baranov,Menotti}. The qualitative difference between
the results obtained within the two approaches is that the velocity
field obtained in \Sec{linres} has irrotational and rotational
contributions at all temperatures, whereas the transport approach
presented in \Sec{transport} gives a purely irrotational velocity
field at zero temperature, as it is the case in homogeneous systems.
The dependence on $\hbar\omega_\pm/\Delta$ mentioned above is missed
within this approach.

In \Sec{results} we used the density matrices obtained in the
preceding sections for the calculation of the moment of inertia. The
result can qualitatively be understood within the two-fluid model,
which describes the superfluid system as a mixture of a superfluid and
a normal-fluid component. The density of the normal-fluid component is
zero at $T = 0$ and approaches the total density for $T\to T_c$. We
have shown that the transport approach exactly reproduces this
two-fluid model. Somewhat surprisingly, the condition for the
transport approach to be valid turns out to be $\hbar\omega \ll
\Delta_0$, where $\Delta_0$ is the value of the gap at $T = 0$. This
is less restrictive than the condition $\hbar\omega \ll \Delta$, in
particular for temperatures near $T_c$.

Within the transport approach, the moment of inertia increases
smoothly from the irrotational limit at $T = 0$ to the rigid-body
value at $T = T_c$. This is a consequence of the increasing density of
the normal-fluid component of the two-fluid model. If the condition
$\hbar\omega \ll \Delta_0$ is not fulfilled, the behavior is
qualitatively similar, but the moment of inertia is always larger than
it is within the transport approach, because in this case the
rotational contributions to the velocity field are always non-zero due
to the finite-size effects mentioned above. In both cases, the
smoothly increasing moment of inertia as a function of temperature can
be obtained only if the thermal quasiparticle occupation numbers are
properly included in the calculation. It is not sufficient to perform
a zero-temperature calculation and then replace the gap $\Delta$ by
the temperature-dependent gap $\Delta(T)$.

Looking at the size of the error made by neglecting the finite-size
effects, we conclude that for the trapped fermionic atoms, where
$\hbar\omega \lesssim \Delta_0$, the hydrodynamical approach is just
at the limit of its applicability. However, we would like to point out
that there are other physical situations, where $\hbar\omega >
\Delta_0$, and where finite-size corrections are crucial. For example,
the moments of inertia of rotating superfluid nuclei ($T = 0$) have at
least twice the irrotational value \cite{Durand}. Also for the
description of superconducting metallic grains in a weak magnetic
field, corresponding to a perturbation $h_1 = (e/mc) \pv\cdot \Av(\rv)
= (e/mc) B_z L_z$ (if $\Bv$ is parallel to the $z$ axis) and therefore
being formally equivalent to a slow rotation, these corrections might
be important.

The method used in \Sec{linres} for the semiclassical solution of the
RPA in superfluid systems can also be extended to the dynamical case,
i.e., to time-dependent perturbations. In this way collective
excitations of the superfluid system, in particular the change of
their frequencies compared with the normal-fluid phase, can be
described. So far the collective modes in the superfluid phase have
been studied either within the hydrodynamical approach
\cite{Zambelli,Baranov} or quantum-mechanically for the case of
spherical symmetry and moderate numbers of particles
\cite{BruunMottelson}.
%
\begin{acknowledgments}
We thank M. Farine and X. Vi\~nas for valuable comments and
discussions. One of us (M.U.) acknowledges support by the Alexander
von Humboldt foundation (Germany) as a Feodor-Lynen fellow.
\end{acknowledgments}
%
\appendix*
\section{Alternative derivation of the Migdal term}
%
In \Sec{linres} we derived the change of the pairing field,
$\Delta_1(\rv)$, via a gauge transformation. Here we will present
an alternative method, which is more direct, but also somewhat more
difficult. We will solve the original integral equation for $\Delta_1$
which one obtains by inserting \Eq{kappa1} into \Eq{gapeq1}:
\begin{align}
\Delta_1(\rv) &= \Delta_1^\mathit{IB}(\rv)+\Delta_1^M(\rv)
  \nonumber\\
 &= g\int\! \frac{d^3p}{(2\pi\hbar)^3}\,
   [\kappa_1^\mathit{IB}(\rv,\pv) + \kappa_1^M(\rv,\pv)]\,.
\label{inteq}
\end{align}
with
\begin{gather}
\kappa_{1\,n\np}^\mathit{IB} = F^{\kappa h}_{n\np} h_{1\,n\np}\,,\\
\kappa_{1\,n\np}^M = F^{\kappa \Delta}_{n\np} \Delta_{1\,n\np}\,.
\end{gather}
The Wigner transforms of these two contributions to $\kappa_1$ can be
calculated semiclassically as given by \Eq{rho1sc} with $\rhotil_1$
replaced by $\kappa_1^\mathit{IB}$ and $\kappa_1^M$, respectively, and
$F^{\rho h}$ replaced by $F^{\kappa h}$ and $F^{\kappa \Delta}$,
respectively.

The first term in \Eq{inteq}, $\Delta_1^\mathit{IB}$, has already been
evaluated in \Sec{linres} [term proportional to $\Omega$ in
\Eq{delta1tilde}] with the result
\begin{equation}
\Delta_1^\mathit{IB}(\rv) =
  \frac{i g m^2 p_F(\rv) r_x r_y}{8\pi^2 \hbar^2 \Delta}
  \Omega \omega_+ \omega_- (G_+ + G_-)\,.
\label{delta1ibresult}
\end{equation}
Now we turn to the evaluation of the second term, $\Delta_1^M$. The
explicit expression for $F^{\kappa\Delta}(\xi,\xip)$ reads
\begin{multline}
F^{\kappa\Delta}(\xi,\xip) =
  \frac{1-2f(E)}{4E}+\frac{1-2f(\Ep)}{4\Ep}\\
  -\frac{[1-f(E)-f(\Ep)](\xi-\xip)^2}{4E\Ep(E+\Ep)}\\
  -\frac{[f(E)-f(\Ep)](\xi-\xip)^2}{4E\Ep(E-\Ep)}\,,
\label{fkdxixi1}
\end{multline}
which in analogy to $F^{\rho h} (\xi,\xip)$ and $F^{\kappa h}
(\xi,\xip)$ can be approximated by
\begin{equation}
F^{\kappa\Delta}\Big(\xib+\frac{\eps}{2},\xib-\frac{\eps}{2}\Big)
  \approx \frac{1-2f(\Eb)}{2\Eb}
  -\Big(\frac{\eps}{2\Delta}\Big)^{\!2} 
      G\Big(\frac{\eps}{2\Delta}\Big)\delta(\xib)\,,
\end{equation}
with $\Eb = \sqrt{\xib^2+\Delta^2}$. Using this approximation we get
[the arguments of the functions $h(\rv,\pv)$ and
$E(\rv,\pv)$ are omitted for brevity]
\begin{multline}
\Delta_1^M(\rv) = g \int\!\frac{d^3p}{(2\pi\hbar)^3}
  \Big[\frac{1-2f(E)}{2E}\Delta_1(\rv)\\
    -\delta(h)\int\! d\eps \Big(\frac{\eps}{2\Delta}\Big)^{\!2}
      G\Big(\frac{\eps}{2\Delta}\Big)\\
      \times \int\!\frac{dt}{2\pi\hbar}\, e^{-i\eps t/\hbar}
        \Delta_1[\rv^\mathit{cl}(\rv,\pv;t)]\Big]\,.
\label{divergent}
\end{multline}
At this stage the disadvantage of the present method as compared with
the method used in \Sec{linres} becomes obvious, since we encounter a
divergent integral over $d^3p$, whereas in \Sec{linres} all
expressions were finite. This divergence is the same one which also
appears in the gap equation (\ref{gapeqlda}) for the unperturbed gap
in local-density approximation. If we assume that this equation is
regularized in some way, we can use it to get rid of the divergence in
\Eq{divergent}, and we obtain
\begin{multline}
\Delta_1^M(\rv) = \Delta_1(\rv)
  +g\int\!\frac{d^3p}{(2\pi\hbar)^3}\,\delta(h)
  \int\! d\eps \Big(\frac{\eps}{2\Delta}\Big)^{\!2}
    G\Big(\frac{\eps}{2\Delta}\Big)\\
      \times \int\!\frac{dt}{2\pi\hbar}\, e^{-i\eps t/\hbar}
        \Delta_1[\rv^\mathit{cl}(\rv,\pv;t)]\,.
\label{delta1m}
\end{multline}
As we will see, the integral equation (\ref{inteq}) can be solved by
the ansatz (\ref{delta1ansatz}). With this ansatz the Fourier
transform in \Eq{delta1m} can easily be evaluated and we obtain
\begin{equation}
\Delta_1^M(\rv) = \Delta_1(\rv)
  +\frac{i g m^2 p_F(\rv) r_x r_y}{8\pi^2 \hbar^2 \Delta}
  \alpha (\omega_+^2 G_+ + \omega_-^2 G_-)\,.
\label{delta1mresult}
\end{equation}
The coefficient $\alpha$ can now be determined by inserting
\Eqs{delta1ibresult} and (\ref{delta1mresult}) into \Eq{inteq}. The
solution, of course, coincides with \Eq{alphasol}.

However, we have to admit that the above arguments concerning the
divergence in \Eq{divergent} are a little bit hand-waving. For
example, \Eq{gapeqlda} (including an appropriate regularization) is
valid only in the local-density approximation, and it does not allow
for a constant gap $\Delta$, while we have for simplicity assumed that
$\Delta$ is a constant in order to derive \Eq{divergent}. Such
inconsistencies do not appear within the formalism presented in
\Sec{linres}.

It remains to show that the Migdal term, calculated as the second term
of \Eq{rho1}, is consistent with the result given in \Eq{rho1mfinal}. This
can be done with the aid of the explicit expression for $F^{\rho\Delta}$,
which turns out to be
\begin{equation}
F^{\rho\Delta}(\xi,\xip) = -F^{\kappa h}(\xi,\xip)\,,
\end{equation}
and the Fourier transform of
$\Delta_1[\rv^\mathit{cl}(\rv,\pv;t)]$.
%
%

\end{document}